\documentclass[9pt,twocolumn,twoside]{pnas-new}
\templatetype{pnasresearcharticle} % Choose template 
\usepackage{pdfpages} 
\makeatletter
\AtBeginDocument{\let\LS@rot\@undefined}
\makeatother
\usepackage{amsmath}
\usepackage{times,mathptmx,amssymb}
\usepackage{graphicx}
\usepackage{color}
\usepackage{siunitx}
\usepackage[english]{babel}
\usepackage{bm}

\usepackage[T1]{fontenc}

\newcommand{\VEC}[1]{\mathbf{#1}}

\title{Hyperuniform states generated by a critical friction field}

\author[a]{Gustavo Castillo}
\author[b]{Nicol\'as Mujica}
\author[b]{N\'estor Sep\'ulveda}
\author[b]{Juan Carlos Sobarzo}
\author[c]{Marcelo Guzm\'an}
\author[b,1]{Rodrigo Soto}
\affil[a]{Instituto de Ciencias de la Ingenier\'ia, Universidad de O'Higgins, Rancagua, Chile}
\affil[b]{Departamento de F\'isica, Facultad de Ciencias F\'isicas y Matem\'aticas, Universidad de Chile, Santiago, Chile}
\affil[c]{Universit\'e de Lyon, ENS de Lyon, Universit\'e Claud Bernard Lyon 1, CNRS, Laboratoire de Physique, Lyon, France}
\leadauthor{Castillo}

\significancestatement{
Hyperuniform systems exhibit random, liquid-like properties at short distances, but reveal an otherwise hidden order at larger scales.  Although many studies exist on static or quasistatic hyperuniformity, little is known about this exotic state of matter under dynamic conditions. Here we observe the emergence of dynamic hyperuniformity during the solid-to-liquid transition in a 2D vibrated layer of granular material.  By characterizing the particle-position fluctuations in Fourier space, we find the robust signature of hyperuniformity on approach to the transition.   With a distilled numerical model that couples the particle density field to a fluctuating frictional field, we fully explain our observations.  Our results provide a first glimpse into hyperuniformity in dynamic systems and a physically intuitive picture for how they arise.
}

\authorcontributions{
N.M and G.C. conceived and planned the experiments. G.C. and J.C.S. performed the experiments. G.C., J.C.S. and N.M. analyzed the experimental data. R.S. and N.S. conceived the theoretical model. N.S. and M.G. ran the numerical simulations. R.S., N.S. and M.G. analyzed the numerical data. All the authors took part in regular discussions and contributed to the writing of the manuscript.}

\authordeclaration{The authors declare no competing financial interests.}

\correspondingauthor{\textsuperscript{1}To whom correspondence should be addressed. E-mail: rsoto@dfi.uchile.cl}


\begin{abstract}
Hyperuniform states are an efficient way to fill up space for disordered systems. In these states the particle distribution is disordered at the short scale but becomes increasingly uniform when looked at large scales. Hyperuniformity appears in several systems, in static or quasistatic regimes as well as close to transitions to  absorbing states. Here, we show that a vibrated granular layer, at the critical point of the liquid-to-solid transition, displays dynamic hyperuniformity. Prior to the transition, patches of the solid phase form, with length scales and mean lifetimes that diverge critically at the transition point. When reducing the wavenumber, density fluctuations encounter increasingly more patches that block their propagation, resulting in a  static structure factor that tends to zero for small wavenumbers at the critical point, which is a signature of hyperuniformity. A simple model demonstrates that this coupling of a density field to a highly fluctuating scalar friction field gives rise to dynamic hyperuniform states. Finally, we show that the structure factor detects better the emergence of hyperuniformity, compared to the particle number variance. 
\end{abstract}

\begin{document}

\maketitle
\thispagestyle{firststyle}
\ifthenelse{\boolean{shortarticle}}{\ifthenelse{\boolean{singlecolumn}}{\abscontentformatted}{\abscontent}}{}

\dropcap{R}ecently, hyperuniform  systems have been identified as an efficient way to fill up space for disordered configurations. These exotic particle distributions are disordered at short distances, as liquids, and more and more uniform when looked at large scales, just as regular, ordered lattices. Hyperuniform states  have been observed in jammed granular~\cite{Donev2005,Berthier2011} and colloidal packings~\cite{Dreyfus2015}, in block-copolymer assemblies~\cite{Zito2015}, in quasicrystals~\cite{quasicrystals2017}, active circle swimmers~\cite{lei2019nonequilibrium}, and even in the patterns of photoreceptive cells in chicken eyes~\cite{Jiao2014}. Recently, these states have been obtained in systems showing non-equilibrium transitions to absorbing states, were hyperuniformity is observed both in the absorbing and fluid phases, close to the transition~\cite{Hexner2015,Weijs2015,hexner2017noise,Hexner4294}. For a recent review, see Ref.~\cite{torquato2018hyperuniform}.
Here, we show that a vibrated granular layer, when approaching the liquid-to-solid transition from the liquid phase, displays  hyperuniform states, which are dynamically generated. A simple model demonstrates that it results from coupling the density to a highly fluctuating  friction field.

 It is possible to characterize the decay of particle correlations  by measuring the average number of particles $\langle N\rangle$ and its variance $\sigma_N^2\equiv\langle N^2\rangle-\langle N\rangle^2$ in boxes of different sizes. In condensed matter, under normal conditions, correlations decay rapidly and  above a certain length, $\sigma_N^2\propto\langle N\rangle$. This is not always the case. For example, in regular lattices, fluctuations take place only at the boundaries. But  also in disordered systems it has been reported that for large sizes $\sigma_N^2\propto\langle N\rangle^{\beta/2}$, with $\beta\neq 2$. When $\beta>2$, the system presents giant density fluctuations, which have been observed in several dynamic non-equilibrium systems as in vibrated nematic granular layers \cite{narayan2007long} or in active matter swarms \cite{Zhang13626}. The opposite case, $\beta<2$, corresponds to hyperuniformity. As the system grows, number fluctuations increase slower than the usual linear behavior, $\sigma_N^2\propto\langle N\rangle$, and density fluctuations become suppressed at the very large wavelength limit. Indeed, another way to characterize particle spatial distributions is via the static structure factor, $S(\VEC{k})\equiv\left(\langle |\widetilde \rho_{\VEC{k}}|^2\rangle - |\langle \widetilde\rho_{\VEC{k}}\rangle|^2\right)/N$, which quantifies the amplitude of the density fluctuations in Fourier space~\cite{chaikin1995principles,Ortiz}. If for small wavevectors it follows a power law $S(\VEC{k})\sim k^\alpha$,  then $\beta=2-\alpha$ for $0\leq\alpha<1$ and $\beta=1$ for $1<\alpha$. Hence hyperuniformity is obtained when $\alpha>0$, that is, when density fluctuations become smaller when reducing the wavenumber  $k$ \cite{Torquato2016}. In other systems, showing stealthy  hyperuniformity,  $S(\VEC{k})$ vanishes in a whole range of small wavevectors~\cite{batten2008classical,florescu2009designer}.
Here we show that, for finite systems, the number variance can give misleading information to detect hyperuniformity. On the contrary, the structure factor works well on finite systems, even in the presence of boundary-induced inhomogeneities, and allows to determine the critical value of the control parameters when hyperuniformity is attained.

\begin{figure*}[t!]
{\includegraphics[width=0.68\columnwidth]{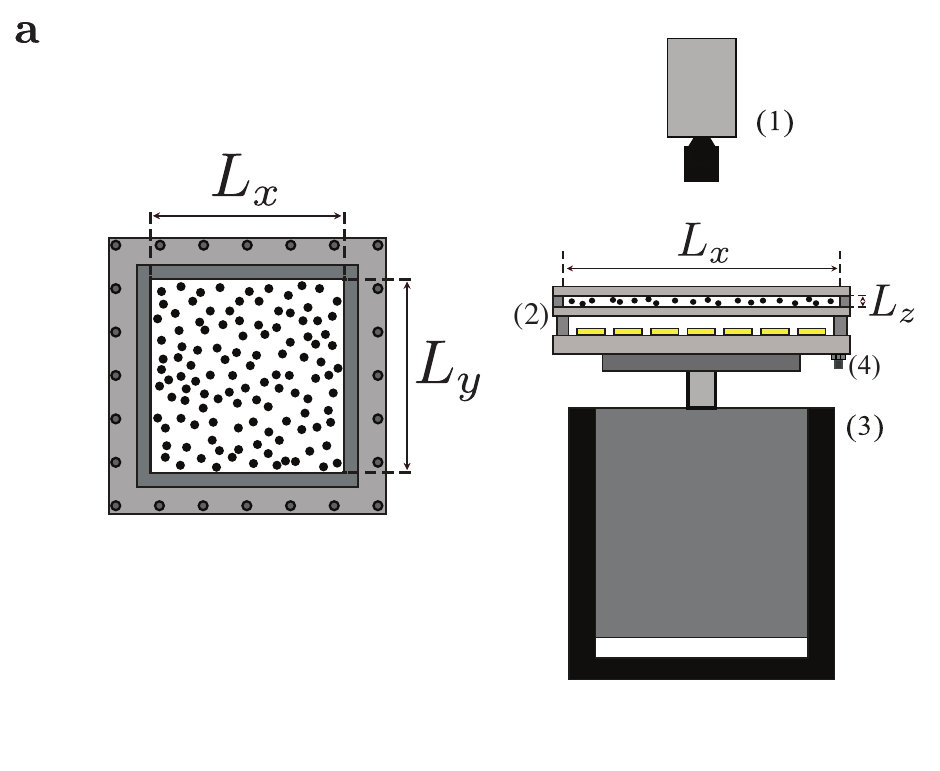}}
{\includegraphics[width=0.68\columnwidth]{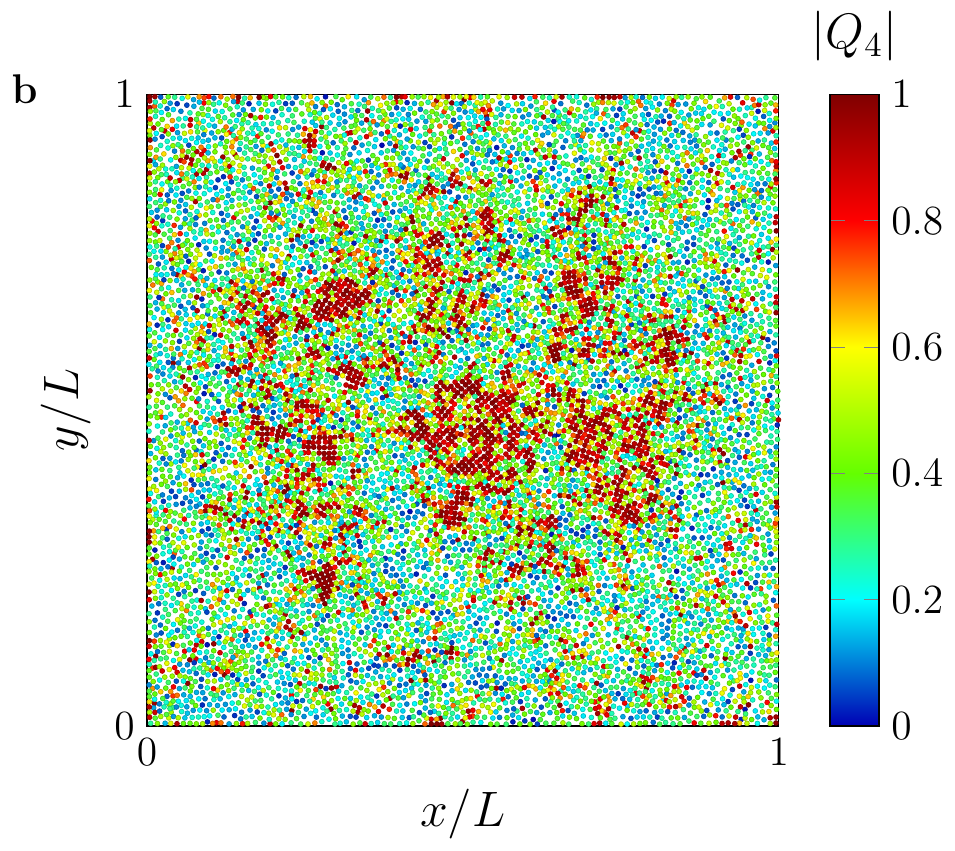}}
\raisebox{0.17\height}{\includegraphics[width=0.515\columnwidth]{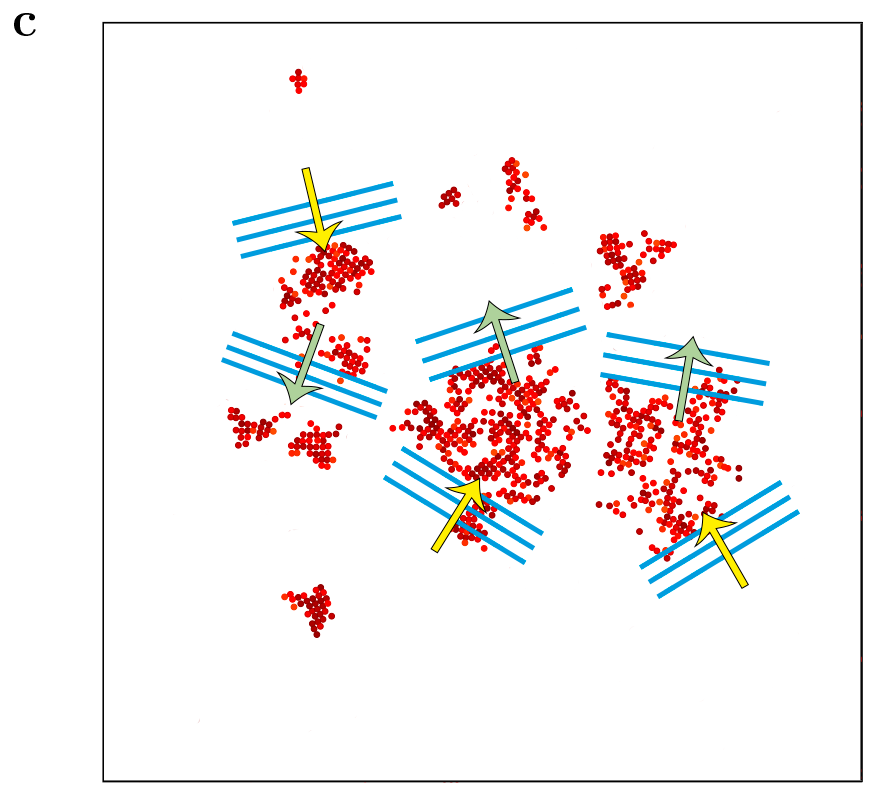}}
\caption{{\bf a}, Schematic of the experimental setup. (left) Top view of the quasi-2D cell, with $L_x = L_y = 100a$. (right) Side view of the setup. The vertical height in the cell is $L_z = (1.94\pm 0.02)a$. The cell is illuminated from below with a 2D array of light emitting diodes, where light is diffused with a white acrylic sheet. (1) camera, (2) quasi-2D cell, (3) electromechanical shaker, (4)~accelerometer. {\bf b}, Color map of the absolute value of the fourfold bond-orientational order parameter in real space, $|Q_4^j|$, for $\Gamma=4.5$. For a better visualization, particles are plotted at 85\% of their diameter. Raw image in Fig. S1. {\bf c},~Schematic representation of the density wave interactions with the crystalline patches shown in {\bf b}.}
\label{fig1}
\end{figure*}

\section{Experiments and simulations of the granular system}

The experimental setup where we observe dynamical hyperuniformity  is the same one reported previously \cite{castillo_prl,Castillo:2015eb}. 
It consists of a rigid shallow box, where $N = 11704$ stainless steel monodisperse spheres are placed inside, with diameter $a = \SI{1}{\milli\meter}$. The box transverse dimensions are $L_x=L_y=L=100a$. Its height is $L_z = 1.94a \pm 0.02a$, such that the  projected 2D filling fraction is $\phi = N\pi a^2/4L^2=0.919$. 
The whole setup is forced sinusoidally with an electromechanical shaker, with vertical displacement $z(t) = A \sin(\omega t)$ (Fig.~\ref{fig1}a). Top view images are obtained with a high-speed camera. The frame rate is either $10$ or $500$ fps, depending on the quantity to be measured. The control parameter is the dimensionless acceleration $\Gamma = A\omega^2/g$, which is varied in the range $1$-$6$, where $\omega = 2\pi f$ is the angular frequency and $g$ the gravitational acceleration. 
See Methods for additional details.

 The box height $L_z<2a$ allows an efficient vertical to horizontal collisional energy transfer mechanism, producing liquid-like states but, at the same time, avoids particles to jump over another. We can then track all particles in the projected two-dimensional motion and therefore analyze simultaneously the microscopic and global dynamics. One of the most remarkable features of this  system is the existence of a liquid-to-solid transition~\cite{Prevost2004,Melby2005}.  When the vibration amplitude increases above a certain threshold, stable crystalline clusters are formed, coexisting with a liquid phase. Notably, this transition can be either continuous or discontinuous depending on the box height. For the former, before the transition, patches of the solid phase form, with length scales and mean lifetimes that diverge critically at the transition point (Fig.~\ref{fig1}b)~\cite{castillo_prl,Castillo:2015eb,Guzman2018}.

\begin{figure*}[t!]
\includegraphics[width=0.68\columnwidth]{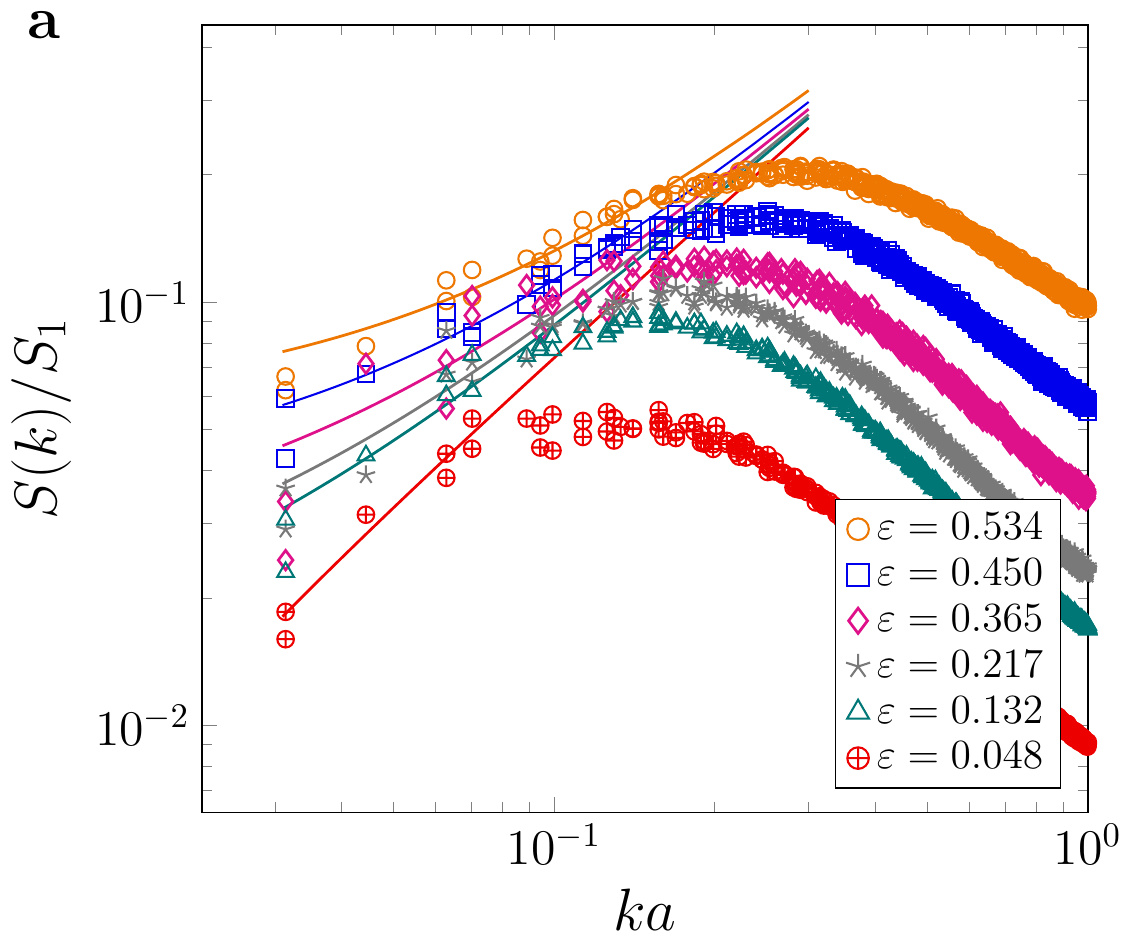}
\includegraphics[width=0.64\columnwidth]{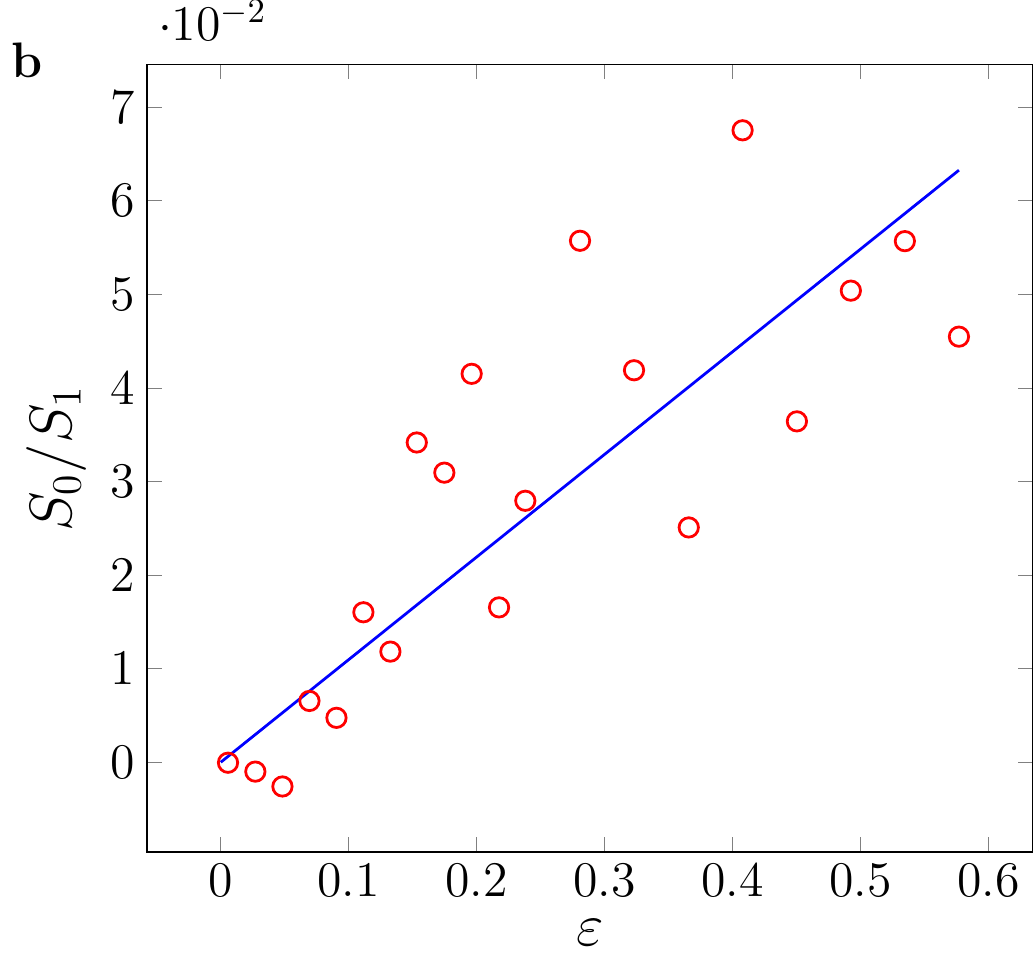} %%
\raisebox{-0.035\height}{\includegraphics[width=0.68\columnwidth]{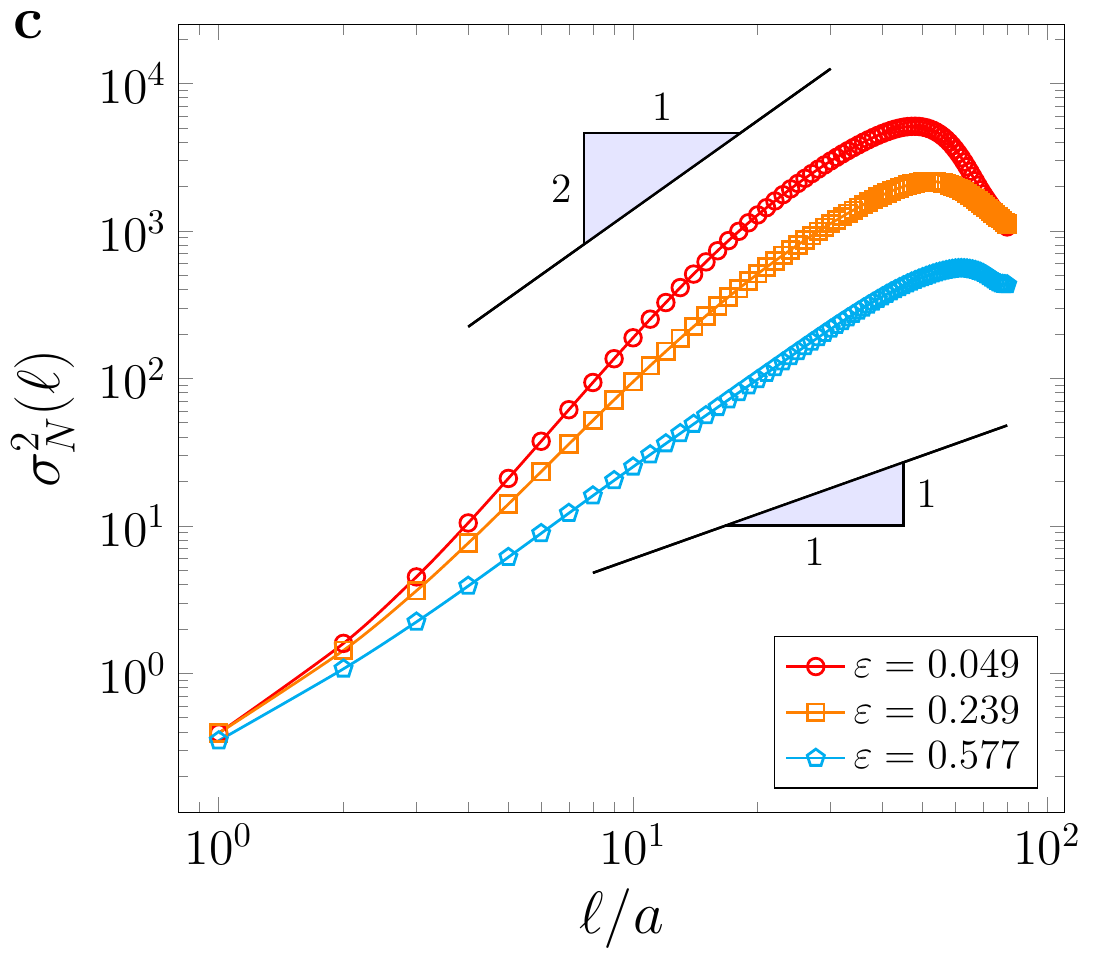}} %%
\raisebox{-0.0\height}{\includegraphics[width=0.68\columnwidth]{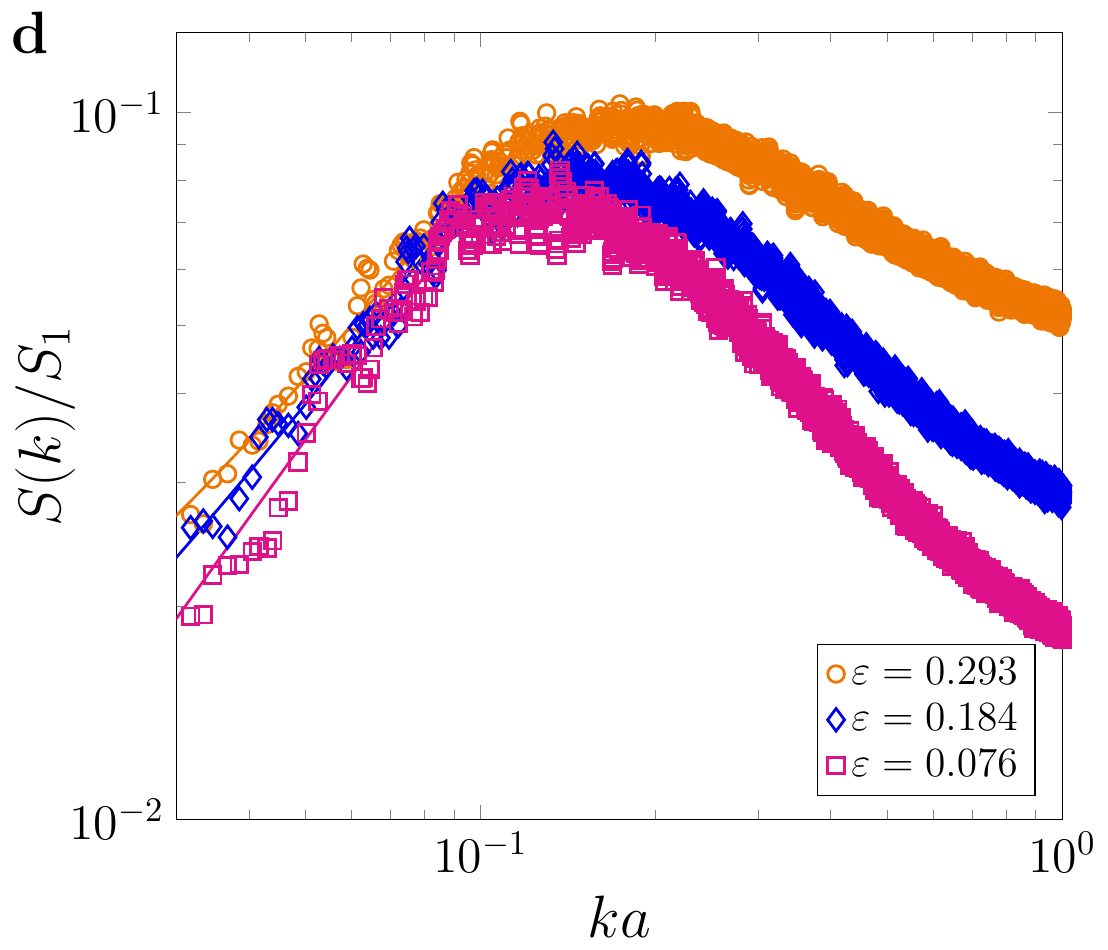}}
\includegraphics[width=0.67\columnwidth]{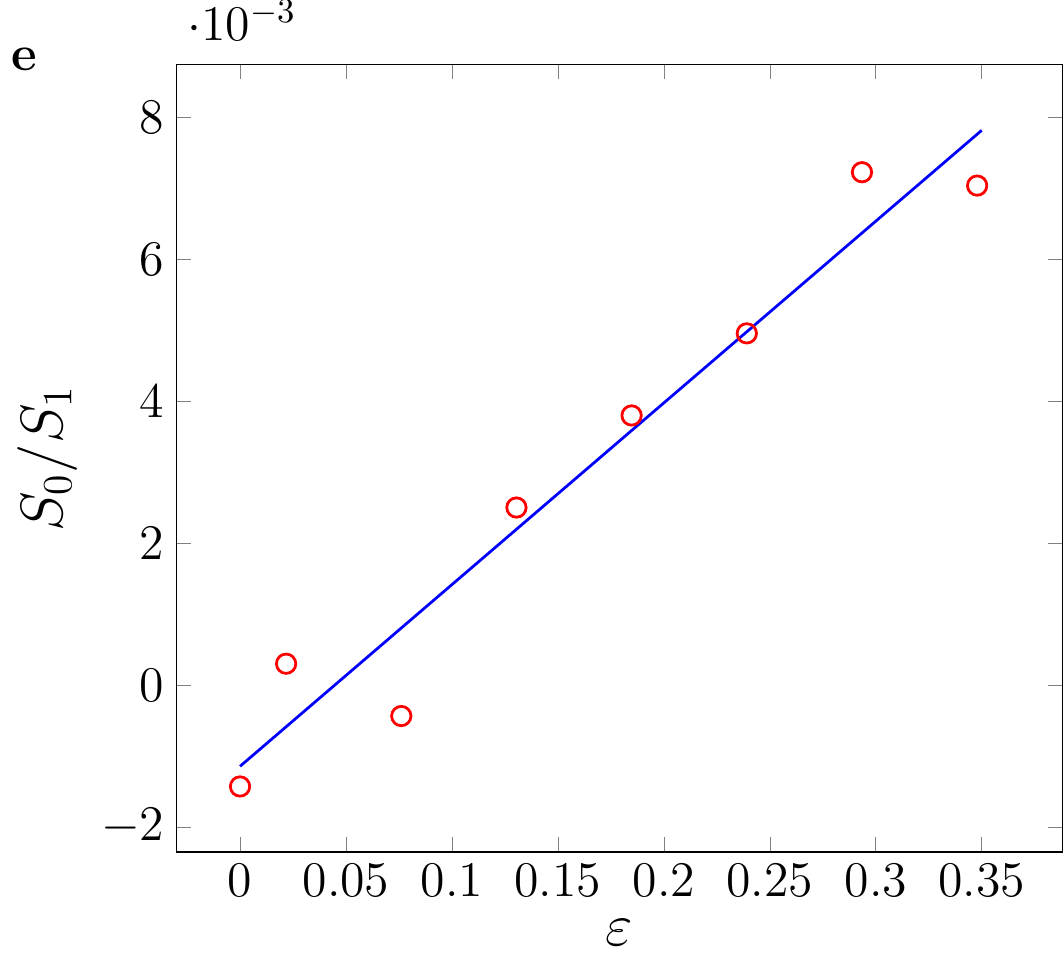} %%
\raisebox{-0.04\height}{\includegraphics[width=0.68\columnwidth]{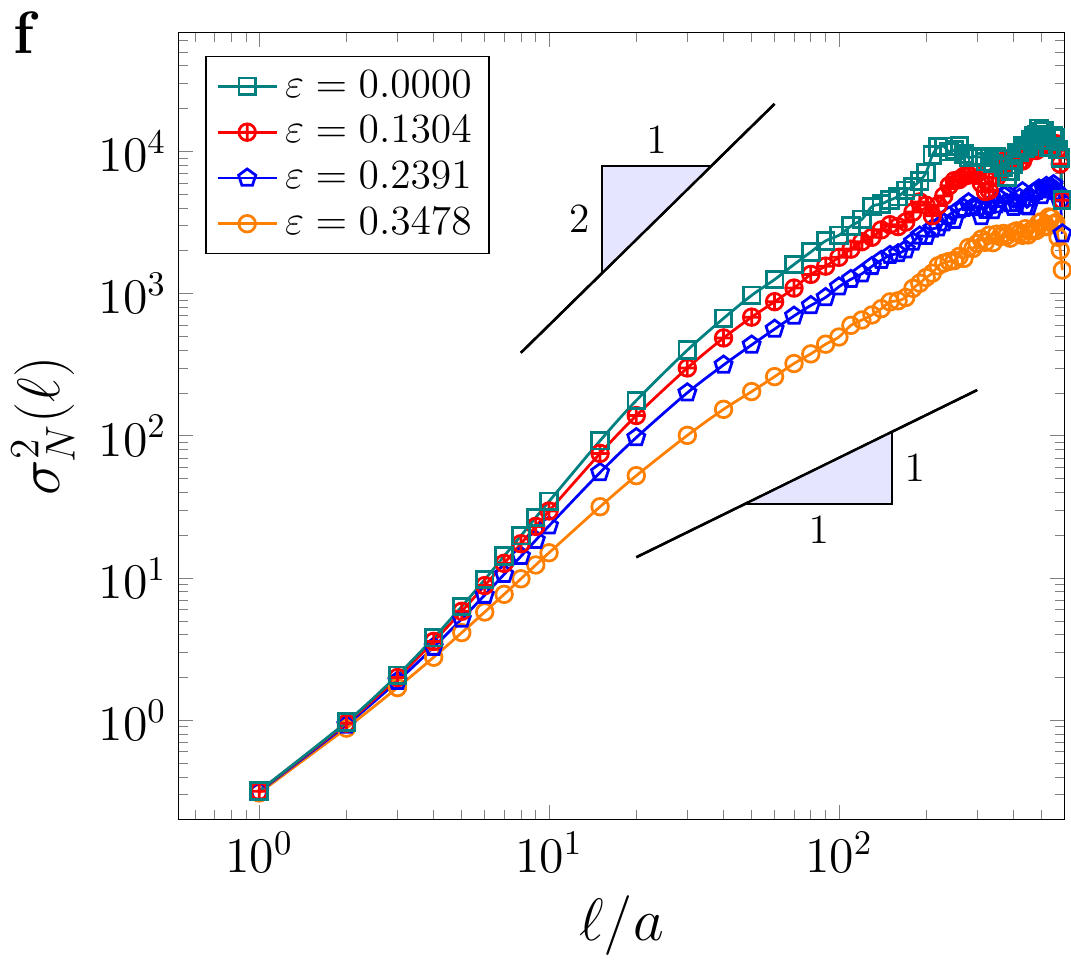}} %%
\includegraphics[width=0.68\columnwidth]{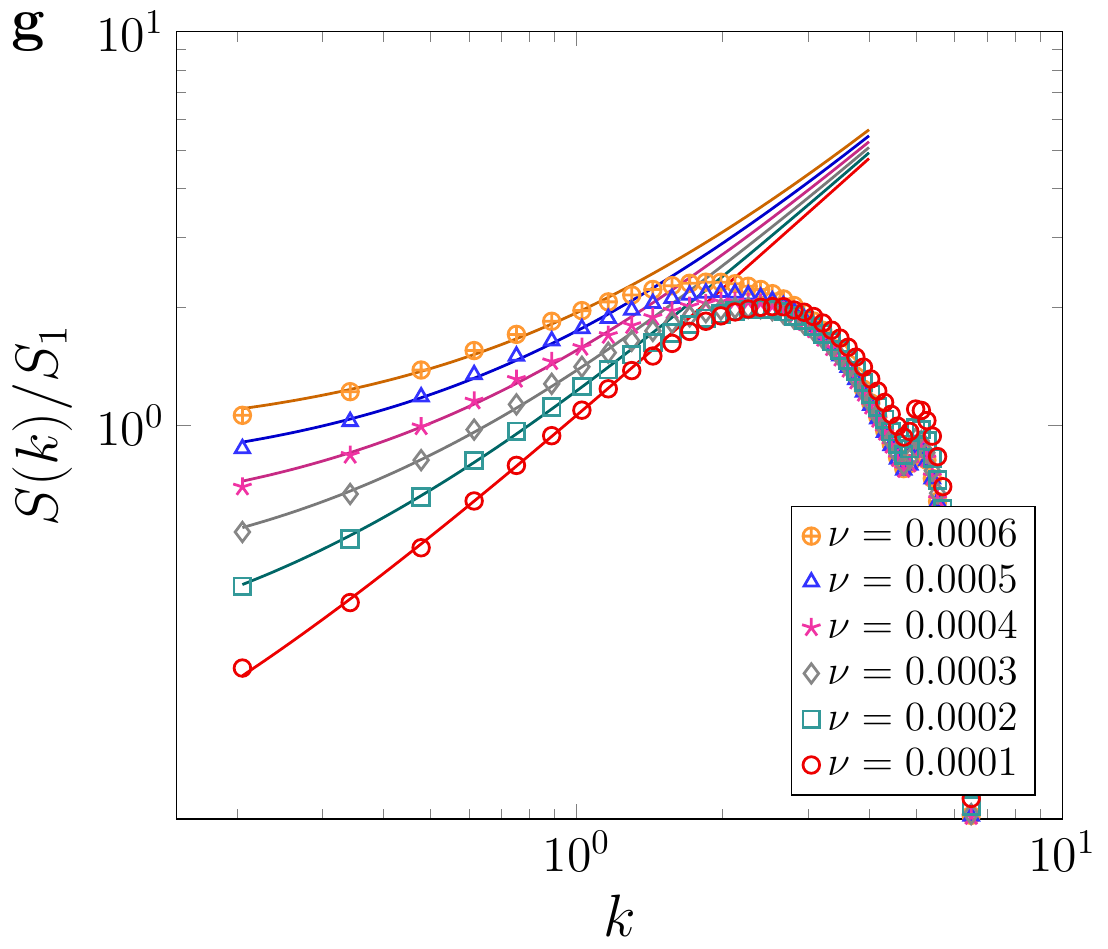}
\includegraphics[width=0.68\columnwidth]{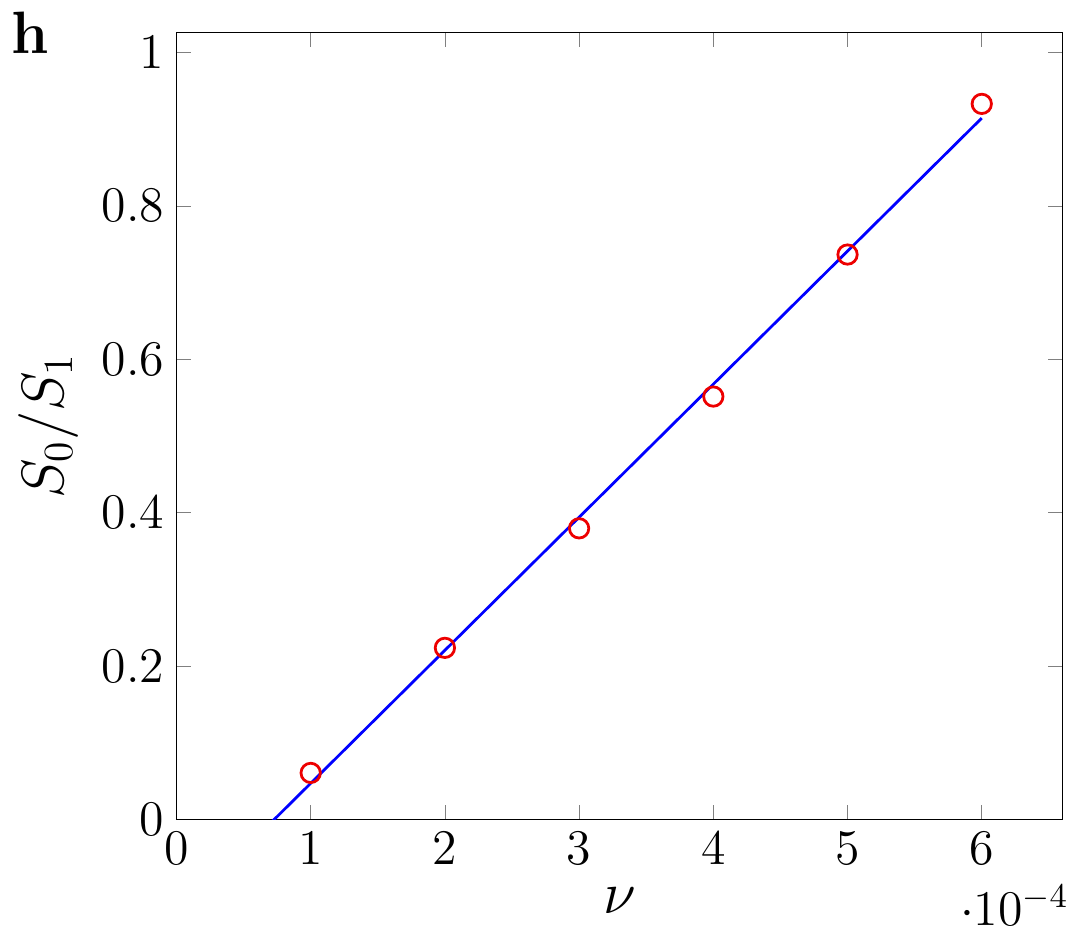} %%
\raisebox{-0.02\height}{\includegraphics[width=0.68\columnwidth]{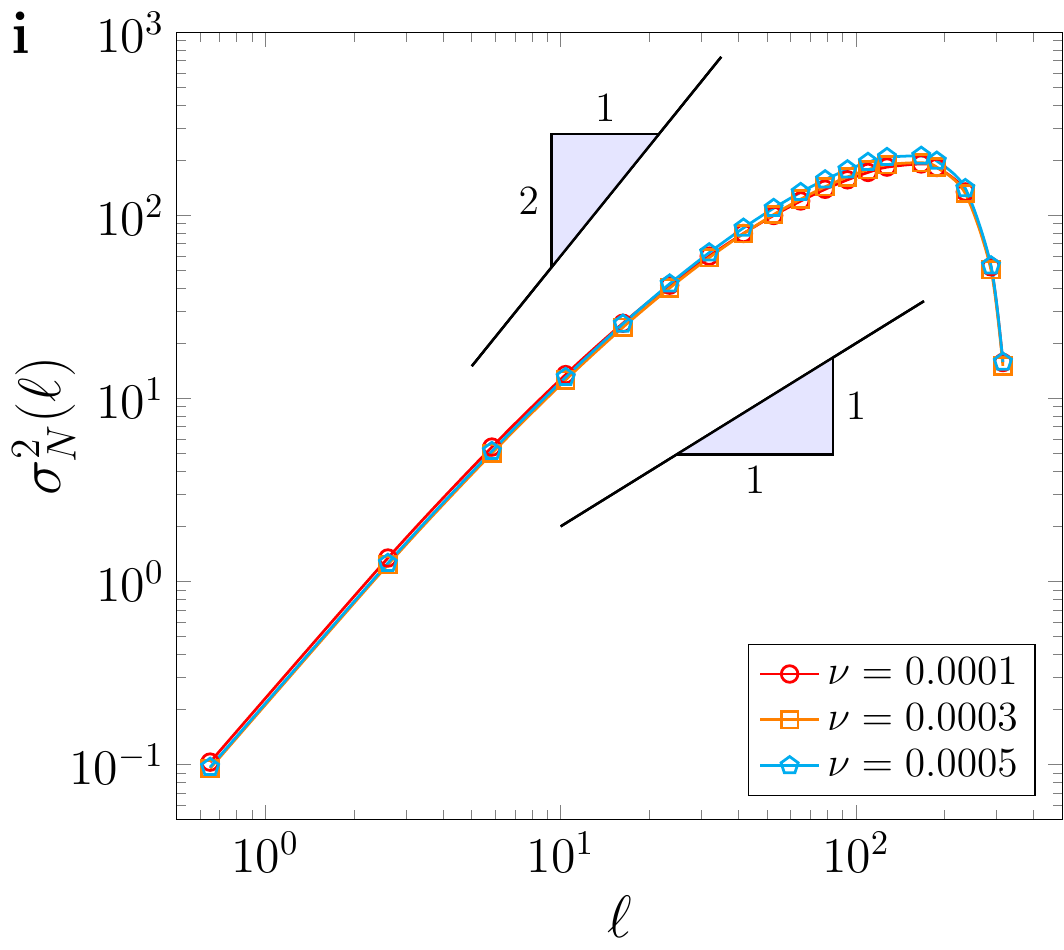}} %%
\caption{
{\bf a}, Experimental static structure factor, $S(k)$, scaled with the slope $S_1$ of the fit $S(k)=S_0+S_1(ka)^\alpha$, with $\alpha=1.12$, in the large wavelength limit for different  reduced accelerations, $\varepsilon\equiv(\Gamma_c-\Gamma)/\Gamma_c$. The solid lines are the results of the fits performed in the range $ka = [0,0.65 k_\text{pp}]$, where $k_\text{pp}$ is the position of the pre-peak (maximum in $S(k)$). 
{\bf b}, Scaled offset $S_0/S_1$ as a function of $\varepsilon$. The solid line is a linear fit passing through zero. 
{\bf c},~Experimental number particle variance $\sigma_N^2$ as a function of the box size $\ell$ for three different $\Gamma$. Note that in the available experimental range, $\sigma_N^2$ does not follow the classical exponent $\beta=2$, neither the hyperuniform exponent $\beta=1$. 
{\bf d}, {\bf e}, and {\bf f}, same as the first row, from the molecular dynamic simulations.
{\bf g}, {\bf h}, and {\bf i}, same as the first row, from  the from  the numerical solution of the model  when $b=4$ in 2D for different values of $\nu$. The fits are $S(k)=S_0+S_1k^\alpha$, for small wavevectors. In {\bf e} and  {\bf h}, the intercept of the linear fit is let free. 
}
\label{fig2}
\end{figure*}

Experiments are performed at fixed vibration frequency, $f = \SI{80}{\hertz}$, and for a box height where the transition is continuous. For these parameters, the critical acceleration is $\Gamma_c = 4.73 \pm 0.15$. The vibration amplitude $A$ is varied in a range below the transition, where the system remains in a stable fluid phase and the solid patches have finite lifetime. Particle positions are obtained from the images, from which it is direct to obtain the two-dimensional structure factor. 
For intermediate wavelengths $ka\sim0.2$, a pre-peak is observed with a height that grows when approaching the transition (see Fig.~S1), although no critical divergence is observed \cite{castillo_prl}. For smaller wavenumbers, the structure factor decreases when $k\to0$ and it can be fitted to the expression  $S(\VEC{k})=S_0 + S_1 (ka)^{\alpha}$ (see Fig.~\ref{fig2}a), where we fix the exponent to $\alpha=1.12$, according to the theoretical model described below. Consistent with the increase in height of the pre-peak, the slope $S_1$ also increases when approaching the transition. Notably, the scaled offset $S_0/S_1$, shown in Fig.~\ref{fig2}b, vanishes at the critical acceleration, resulting in a hyperuniform state.

In order to analyze larger systems, we employ molecular dynamic simulations using the frictional inelastic hard sphere model \cite{marin1993efficient,poschel2005computational} with identical spherical grains and periodic boundary conditions.
We ran the simulations for $N=355500$ particles in a $600a \times 600a \times 1.84a$ system. The parameters are chosen to present the same critical liquid-to-solid phase transition as in the experiments, being therefore suitable to analyze the existence of hyperuniformity (see Methods for details).
In this case, thanks to the use of periodic boundary conditions, statistically homogeneous configurations are obtained and, hence, the structure factor is 
simply computed as $S(\VEC{k})\equiv \langle |\widetilde \rho_{\VEC{k}}|^2\rangle/N$. 
The structure factor can be fitted to the same expression used for the experimental results, and the scaled offset vanishes also at the critical acceleration (see Figs.~\ref{fig2}d and e).

\section{Particle number variance}

\begin{figure}[t!]
\hbox{\hspace{+0.34em} \includegraphics[width=0.85\columnwidth]{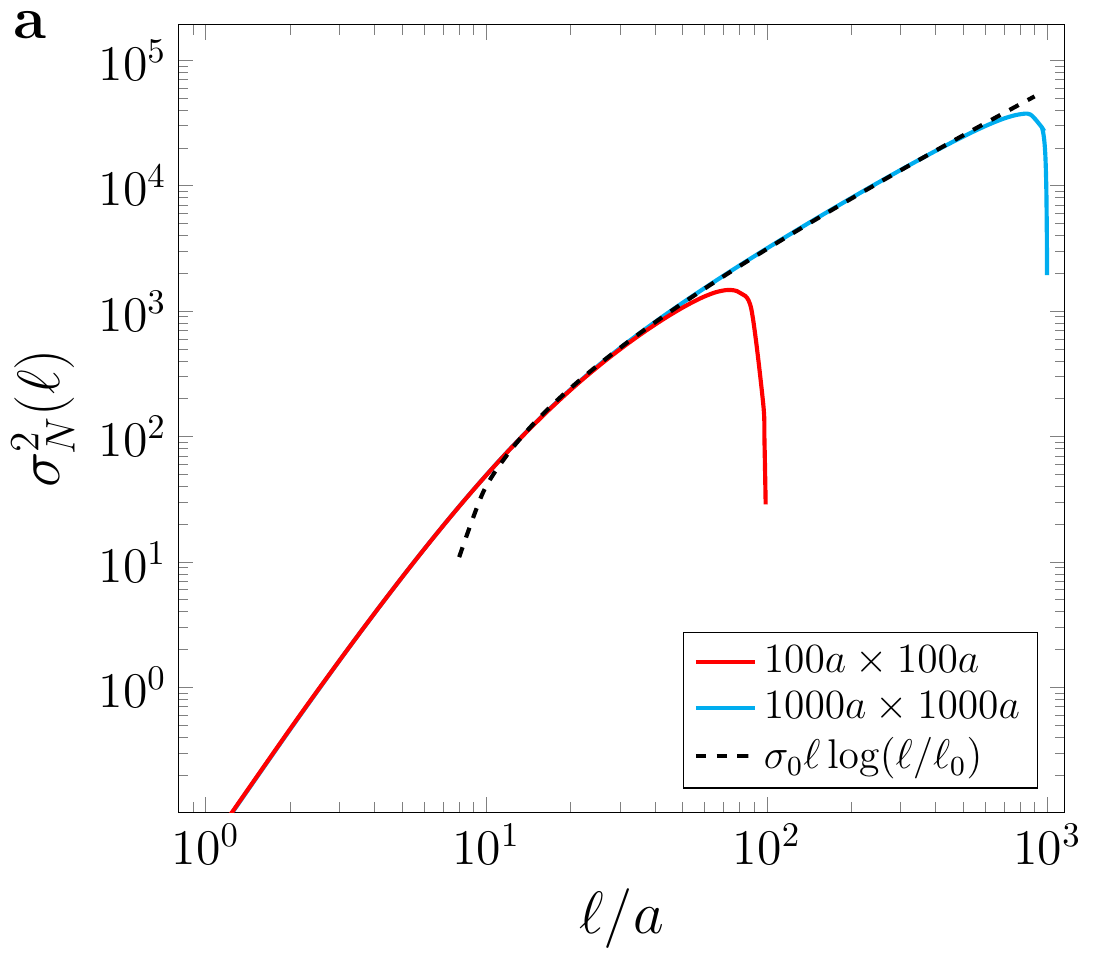}}%%
\includegraphics[width=0.85\columnwidth]{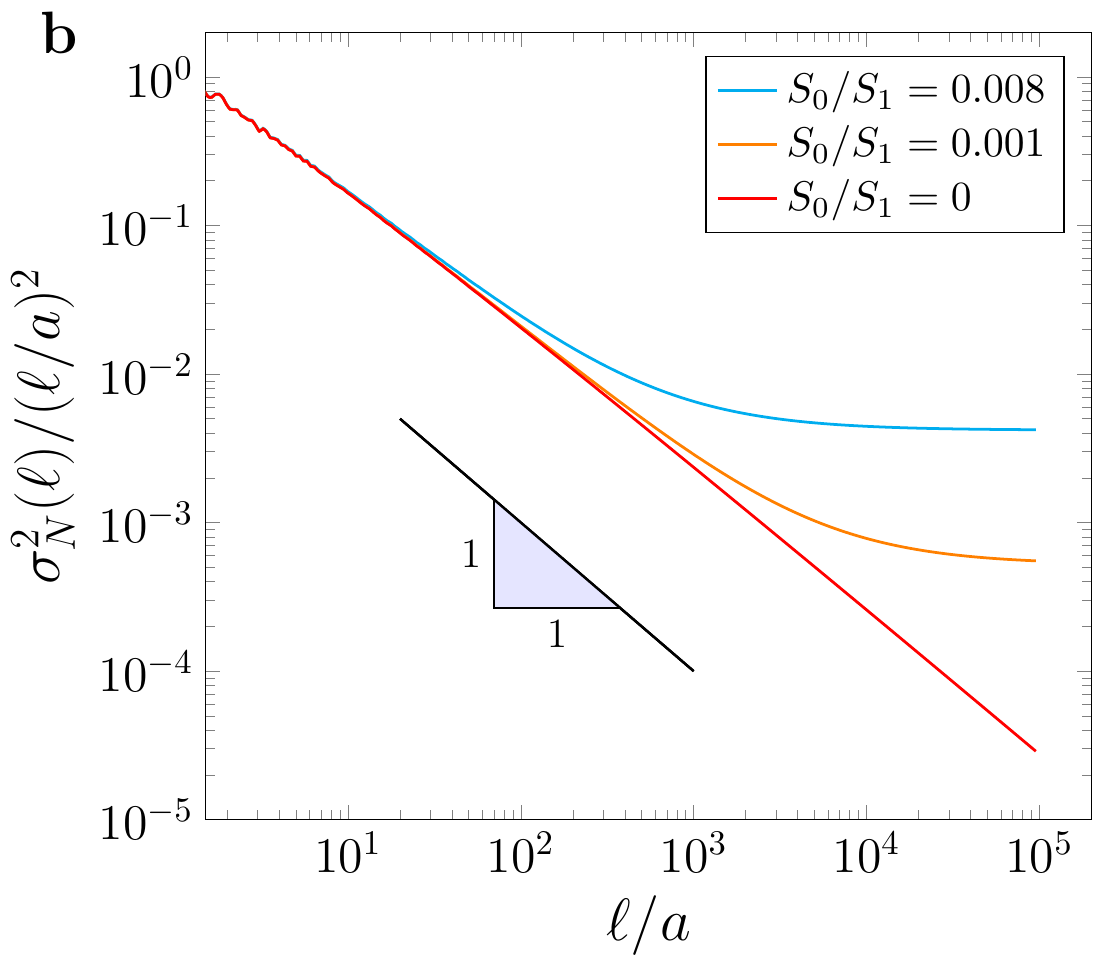} %%
\caption{
{\bf a},~Number particle variance $\sigma_N^2$ obtained from synthetic data using the experimental values of $S(k)$, for two different system sizes $L$. The dashed line shows the expected law for $\alpha=1$. {\bf b},~Normalized number particle variance $\sigma_N^2/\ell^2$ in the thermodynamic limit, $L\to\infty$, obtained from synthetic data using the simulation values of $S(k)$, for different values of $S_0/S_1$.  See SI Text for details.
}
\label{figvariance}
\end{figure}

For boxes of lateral sizes $\ell$, a typical landmark of hyperuniformity is the existence of a sublinear growth of the particle number variance with  $\ell^2$.   
Figure \ref{fig2}c shows the results of the experiments (see Methods), which contrary to the what is displayed by the structure factor, does not present any clear tendency. In fact, the variance looks closer to $\ell^2$ as for a non-hyperuniform material. This apparent contradiction results from the use of a finite system. Indeed, to observe any effect on $\sigma^2_N$, we need to have a full scale separation between the system size and the wavelength above which the structure factor displays the power law $S(\VEC k)\sim k^\alpha$. In our experiments $L=100 a$, with a structure factor peak  located at $ka\sim 0.2$, leaving a limited range to observe the sublinear behavior (see Fig.~\ref{fig2}a). Furthermore, by construction, the variance must vanish for a box equal to the system size, limiting even more the range of box sizes where hyperuniformity can be observed.  
Using the relation between the structure factor and the particle number variance~\cite{PhysRevE.68.041113}, we generate 
synthetic values of $\sigma_N^2$ using the experimental values of $S(k)$ for other system sizes (see SI Text for details). It is found that the sublinear growth in $\sigma_N^2$ would only be observed for large system sizes, $L\gtrsim1000a$ (Fig.~\ref{figvariance}a). 

In the case of the simulations, where the system is larger, the sublinear growth of $\sigma_N^2$ is indeed observed (Fig.~\ref{fig2}f). However, to our surprise, the hyperuniform behavior appears also for finite distances to the critical point, for which the offset $S_0/S_1$ does nor vanish. Figure \ref{figvariance}b presents $\sigma_N^2/\ell^2$ with  synthetic generated values of the number variance using the simulation form of $S(k)$ at different distances to the critical point. As expected, at the critical point $\sigma_N^2/\ell^2\sim\ell^{-1}$. At finite distances to the critical point, $\sigma_N^2/\ell^2$ first decays with $\ell$ to finally attain a constant value, consistent with the finite value of  $S_0/S_1$. Notably, the crossover $\ell^*\approx 2\pi a(S_0/S_1)^{-1/\alpha}$ between the two regimes, takes place at quite large values of $\ell$, which can be larger than $L$. Hence, for finite system sizes, the decay in  $\sigma_N^2/\ell^2$ can give the erroneous indication that the system presents hyperuniformity. On the contrary, the analysis of $S(k)$ is robust to detect the critical value of the control parameters where hyperuniformity is attained. 

\section{Model}
\begin{figure}[t!]
\includegraphics[width=0.95\columnwidth]{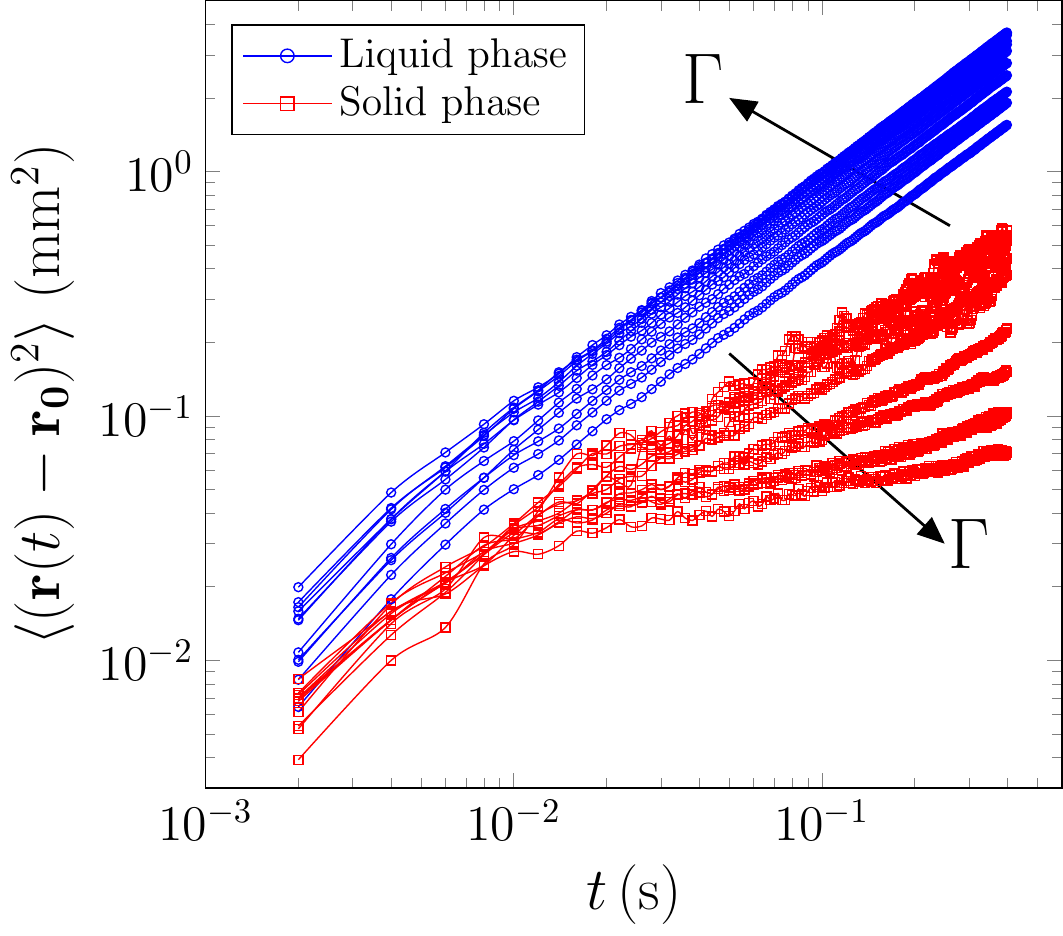}
\caption{Measured mean-square displacement versus time for $\Gamma \in [2.0,5.8]$. Liquid particles obey a diffusive behavior, with a diffusion constant that increases with acceleration; on the contrary, solid particles are subdiffusive, presenting an increasing amount of subdiffusivity for increasing $\Gamma$.}
\label{fig3}
\end{figure}

The observed hyperuniformity is closely linked to the existence of the intermittent solid patches near the critical point, as we argue here. 
In Ref.~\cite{castillo_prl} we demonstrate that the relevant fields to describe the system close to the transition  are the particle density $\rho$, which is conserved, and the four-fold order parameter $Q_4$, which measures the fraction of particles in the solid phase and their degree of order. For each particle, we compute
\begin{equation}
Q_4^j = \frac{1}{N_{j}} \sum_{s = 1}^{N_{j}} e^{4i\theta{s}^j},
\end{equation}
where $N_j$ is the number of nearest neighbors of particle $j$ and $\theta_s^j$ is the angle between the neighbor $s$ of particle $j$ and the $x$ axis.  For a particle in a square lattice, $|Q_4^j | = 1$ and the complex phase measures  the square lattice orientation. A map of $|Q_4^j|$ for  $\Gamma= 4.5<\Gamma_c$ is shown in Fig.~\ref{fig1}b. Below the transition, the structure factor of $Q_4$ presents an Ornstein-Zernike form, $S_4(\VEC k)=S_{4}(0)/[1+(\xi_4 k)^2]$, where the amplitude $S_{4}(0)$ and correlation length $\xi_4$ diverge at the critical point. 
Friction with the top and bottom walls is always present implying that the lateral momentum is rapidly dissipated  \cite{Clerc:2008hi}. As a consequence, in the slow large-scale dynamics, particle density evolves diffusively. 
 Through  $Q_4^j$, particles can be classified to belong to the liquid or the solid phase. We have previously shown that the stationary distribution of $|Q_4^j|$ is bimodal, with a wide maximum around $|Q_4^j| = 0.3$, and a much narrower one at $|Q_4^j| \approx 0.95$ (details in the Supplementary Material of Ref.~\cite{PRE_Luu}). The local minimum between the two peaks is at $|Q_4^j| = 0.7$. To determine the diffusive nature of particles in each phase we track them for a period of time $T_o = \SI{0.4}{\second}$,  large compared to the fast time scale of energy injection and dissipation (vibration period $1/f =\SI{12}{\milli \second}$). To compute the mean-square displacement of particles in each phase, we impose that the time average $\langle |Q_4^j| \rangle_{T_o}$, computed for the observation time $T_o$, must be $\leq 0.4$ or $\geq 0.7$ to be classified as liquid or solid particles, respectively (see SI Text). Next, the mean-square displacement of the ensemble of particles that satisfy the previous conditions is computed using the tracked trajectories.
We find that the behavior  is radically different for the liquid and solid phases (see Fig.~\ref{fig3}). The former exhibits diffusion, while the later shows subdiffusive dynamics. This subdiffusion can be modeled by a very small diffusion coefficient, a property that is a result of caging and the enhanced friction from the repeated rapid collisions with the top and bottom walls. The solid patches and the liquid phase do not differ substantially in density but, as a result of order, present  important differences in their diffusive dynamics. 
Note, however, that the system is not frozen in a glassy state as patches form and disappear continuously, and the subdiffusive dynamics is observed only for times comparable to their lifetime.  
Density fluctuations with wavevector $\VEC k$ interact with these patches of reduced mobility, disturbing their propagation. At the critical point, the patches are scale free, with an intensity and lifetime that increases when decreasing $k$~\cite{castillo_prl}, blocking efficiently the density fluctuations at large scales, resulting in an hyperuniform state (Fig.~\ref{fig1}c).  To put this hypothesis into test, we build a simple model which retains the principal exposed features, namely the existence of an order parameter with critical dynamics, which controls the diffusion of the conserved density field. 

With these elements, we propose the model,
\begin{equation}
\frac{\partial \rho}{\partial t}=\nabla\cdot\left(D\nabla \rho + \bm{\eta} \right).
\label{cont_eq_for_n_2}
\end{equation}
Here, $\bm{\eta}$ is a fluctuating mass flux, which is modeled as a  white noise satisfying 
$\langle \VEC \eta(\VEC{r},t) \rangle=0$ and $\langle \eta_i(\VEC{r},t) \eta_k(\VEC{r}',t') \rangle=C_1\delta_{ik}\delta(t-t')\delta(\VEC{r}-\VEC{r}')$, where $i,k = \{1,2\}$ are the Cartesian coordinates. The   diffusion coefficient  $D$ depends on a local order field $\psi$, related to $Q_4$, which is described by the critical equation
\begin{equation}
\frac{\partial \psi}{\partial t} = \mu\nabla^b\psi -\nu\psi + \xi,
\label{diff_eq_for_mu1}
\end{equation}
where $\xi$ is a white noise satisfying
$\langle \xi(\VEC{r},t) \rangle=0$ and $\langle \xi(\VEC{r},t) \xi(\VEC{r}',t') \rangle=C_2\delta(t-t')\delta(\VEC{r}-\VEC{r}')$. The constant $\mu$ accounts for the spatial coupling of $\psi$, while $\nu$ measures the distance to the critical point. The order parameter $\psi$ presents fluctuations that grow when approaching the transition and, hence, we identify large values of $\psi$ as being representative of the solid phase. Then, the diffusion coefficient is modeled as $D=D_0e^{-\lambda \psi^2}$, taking finite values in the liquid phase, while vanishing asymptotically in the solid phase. Other expressions for $D$ with the same asymptotic limits generate similar properties. To fix the  exponent $b$, we require that the $\psi$-$\rho$ coupling becomes relevant, in the renormalization group sense, at large scales.
Considering the scaling $\VEC{r}\rightarrow s \VEC{r}$, $t\rightarrow s^z t$ and $\psi\rightarrow s^{\chi}\psi$, the different parameters scale as  
$\mu\rightarrow s^{z-b}\mu$, $\nu\rightarrow s^z \nu$, and $C_2\rightarrow s^{z-d-2\chi}C_2$, where $d$ is the spatial dimensionality.
Units are fixed by choosing that $\mu$ and $C_2$ are not modified when the spatial scale $s$ is changed, resulting in $z=b$ and $\chi=(b-d)/2$. For the $\psi$-$\rho$ coupling to become relevant, $\psi$ must take large values close to the critical point at increasingly large spatial scales. This is obtained when $\chi>0$ or, equivalently, $b>d$. That means that for $d=1$ and $b=2$ (normal spatial coupling for $\psi$) it is possible to observe large scale structures close to the critical point while for $d=2$, a larger exponent is needed, for example $b=4$ to keep analyticity. The need of a large exponent suggests that this is an effective description of other microscopic fields obeying normal diffusion.  For the purpose of this letter, it is not necessary to investigate further into these eventual underlying dynamics.

The numerical solution of the model (see Methods for details) shows that the structure factor of $\psi$ displays an Ornstein-Zernike form. The density field static structure factor  indeed presents a notorious decrease when reducing the wavevector, which is well fitted as  $S(\VEC k)= S_0+S_1k^{\alpha}$.
Imposing that $\alpha$ is the same for all values of $\nu$, we obtain $\alpha = 1.58\pm0.10$ in 1D, independent of the spatial coupling exponent $b$, and in 2D we get $\alpha = 1.12\pm 0.15$, for $b = 4$ (see Fig.~\ref{fig2}d and SI Text). As in experiments, varying $\nu$ the scaled offset $S_0/S_1$ vanishes and hyperuniformity is obtained (Fig.~\ref{fig2}e). Hyperuniformity appears at a small but finite value of $\nu$, probably due to renormalization of the bare critical point. Within the measurement precision, in experiments and simulations, hyperuniformity takes place at the critical amplitude  (Fig.~\ref{fig2}b).

\section{Discussion and conclusions}
We have shown that hyperuniformity can be dynamically generated in fluid-like states if friction is highly heterogeneous. 
%\com{Lo sacamos? The idea of a fluctuating friction field has been used to describe the flow of granular materials \cite{kamrin2007stochastic}.} 
In our model density evolves by diffusion, with a diffusion coefficient that can become very small in fluctuating patches. In systems where momentum is conserved, a similar behavior should be obtained if the viscosities are heterogeneous, with patches of high dissipation. 

In small systems, the finite size effects make it difficult to identify hyperuniformity with the  particle number variance. On the other hand, in the analysis of density fluctuations the finite size effects, boundary inhomogeneities, and small inhomogeneities in the setup do appear in both $\langle |\widetilde \rho_{\VEC{k}}|^2\rangle$ and  $\langle \widetilde\rho_{\VEC{k}}\rangle$ (for example, there are strong differences between odd and even modes), but notably they cancel out in $S(\VEC{k})$. Indeed, the structure factor, which describes the dynamics of the fluctuations~\cite{chaikin1995principles,Ortiz}, is remarkably isotropic and smooth~\cite{Castillo:2015eb}. This feature makes $S(\VEC{k})$ an ideal observable to identify hyperuniform states.

Hyperuniformity is characterized by positive values of the exponent $\alpha$. Here we report the values $1.12$ (experiment and model in 2D) and $1.58$ (model in 1D). This diversity is consistent with a similar variability reported in the literature by measuring the structure factor: $0.45$ in critical absorbing states~\cite{Hexner2015}, $0.5$ in periodically driven emulsions~\cite{Weijs2015}, $1$ in jammed packings of polydisperse spheres~\cite{Donev2005,Berthier2011}, and $1$ in the photoreceptive cells of chicken eyes~\cite{Jiao2014}, or deducing it from the measurement of the number variance exponent: 0.8 in block-copolymer assemblies~\cite{Zito2015}, 0.21 in jammed packings of soft spheres~\cite{Dreyfus2015}, 1 in two-phase random  media~\cite{Zachary2009}, and 2 in quasicrystals~\cite{quasicrystals2017}. The case of systems with absorbing states are related to the universality class of the Manna model~\cite{Hexner2015,hexner2017noise}, where a field theory analysis suggests that it can interpreted as a field-dependent diffusion coefficient~\cite{pastor2000field}. The field equations, however, are different to the model presented here.
It remains to be understood if this range of values is a signature of  different universality classes or if $\alpha$ is
not an universal exponent at all.

\matmethods{

\subsection*{System setup}
\label{sec_expint}

The box consists of two 10-mm-thick glass plates separated by a square metallic frame. Each inner glass surface has an indium tin oxide (ITO) coating, which dissipates electrostatic charges generated by collisions of particles with the walls. A piezoelectric accelerometer is fixed to the base, allowing the measurement of the imposed forcing acceleration with a resolution of 0.01$g$. For each $\Gamma$, three videos of 3000 images were acquired, two at $10$ fps for the computation of $S(\VEC k)$ and one at $500$ fps to obtain the diffusion coefficients. The acquired images have a resolution of $1600\times1600 \mbox{ pix}^2$. Particle positions are determined at sub-pixel accuracy. The particle detection is done by using a modified open source Matlab code, which uses a least-square algorithm \cite{Shattuck}. Our modified version in C++ and CUDA allows faster computation for a large number of particles \cite{ThesisJuan, PaperNicoScott}. The algorithm allows us to detect both layers of particles in a dense solid cluster, where the top-layer particles are placed in the valleys that the bottom particles form.

\subsection*{Molecular dynamics simulations}

The grain-grain and grain-wall friction coefficients are fixed to $\mu=0.03$, and the restitution coefficient to $\alpha=0.998$, which were chosen by inspection to ensure the liquid-to-solid phase transition \cite{guzman2018critical}. The friction coefficients are one order of magnitude smaller than the experimental values. This artifact is used to map the simulations with perfectly spherical grains and flat walls to the experiments that present small imperfections in roughness and flatness.   While a quantitative comparison with the experiments would require to take into account these details~\cite{rivas2011sudden}, our simulations reveal the same phenomenology as the experiments, namely a critical transition. 
The oscillation frequency is fixed to $\omega= 5 \sqrt{g/a}$, where $g$ is the gravity acceleration, and the oscillation amplitudes where varied in the range $A \in (0.06 a, 0.1 a)$. The critical amplitude was found to be $A_c=0.092 a\pm 0.001$, giving $\Gamma_c=A_c\omega^2/g=2.3$. 

\subsection*{Experimental number particle variance}

In order to quantify the spatial correlations, the average number of particles $\langle N\rangle$ and its variance $\sigma_N^2\equiv\langle N^2\rangle-\langle N\rangle^2$ in boxes of different sizes are measured.  From each image we compute the number of particles in subsystems of different size, defined by square windows ranging in size from $a\times a$ to $80a\times80a$. Due to the presence of some spatial heterogeneities, each square window of size $\ell$ is displaced throughout the entire image (excluding $10a$ at each border), which gives us a spatial average of $N$. Then, for the set of images at a fixed $\Gamma$ we determine, for each subsystem size, the average $\langle N \rangle $ over all images and thus, its standard deviation, $\sigma_N(\ell)$. 
%In Fig.4a of the paper we shown the behavior of $\sigma_N^2(\ell)$ for different accelerations $\Gamma$. 

%\subsection*{Classification of particles in liquid and solid phases for computing the mean square displacement}
%
%In Fig. \ref{fig2-si}a we present two typical time series of $|Q_4^j|$, for a particle in the solid phase and one in the liquid phase for a complete time duration of $0.4$ s. The former has a high average $|Q_4^j|$, whereas the later has a lower average; both realizations show fluctuations. In Fig. \ref{fig2-si}b we present the standard deviation of  $|Q_4^j|$,  $\sigma_{|Q_4^j|}$, versus $\langle |Q_4^j| \rangle$ for $\Gamma = \{2.5,4, 6\}$. At lower and intermediate forcing most particles are in the liquid phase, with some particles that occasionally condense, temporally, in small ordered solid clusters. Well above the transition there are many particles that stay in the solid phase during the complete time series. At intermediate and high accelerations a large fraction of particles fluctuate between both phases randomly, as manifested by the larger measured $\sigma_{|Q_4^j|}$, with a maximum around $|Q_4^j| = 0.55$. Thus, in order to compute a dynamic quantity, such as the mean square displacement, we choose particle trajectories with small fluctuations (lower $\sigma_{|Q_4^j|}$). In practice, to do so, we select particle trajectories with well defined averages: $|Q_4^j|\leq0.4$ for liquid particles and $|Q_4^j|\geq0.7$ for solid ones. 
%

\subsection*{Dimensionless model}

The  parameters of the dimensional model are $b$, $\mu$, $\nu$, $C_1$, $\lambda$, $C_2$ and the system length $L$. The dimensions of $\psi$ are arbitrary then we choose to make it dimensionless. As the product $\lambda \psi^2$ must be also dimensionless, we can take $\lambda=1$. Considering \eqref{diff_eq_for_mu1}, the dimensions of the parameters are   $[\mu]=\frac{L^b}{T}$ and 
$[C_2]=\frac{L^d}{T}$. It is then possible to fix length and time units such that $\mu=C_2=1$. By \eqref{cont_eq_for_n_2},  $[C_1]=\frac{[\rho]^2L^{d+2}}{T}$. As the unities of $\rho$ are arbitrary, it is possible to choose  the units of $\rho$ in such a way that $C_1=1$. With these choices, the dimensionless model has only $b$, $\nu$, and $L$ as free parameters.

\subsection*{Numerical solution of the model}
The  integration of \eqref{diff_eq_for_mu1} was performed using the Crank--Nicolson method~\cite{press2007numerical}, where the spatial part is computed with spectral methods using the  FFTW library~\cite{FFTW05}.  \eqref{cont_eq_for_n_2}  was integrated with the Alternating-Direction Implicit method~\cite{press2007numerical}.
The characteristic time and lengths scales are $t_c=1/\nu$ and $\ell_c=t_c^{1/b}$. Considering the spatial and temporal discretization $\Delta x$ and $\Delta t$, respectively, large structures will be well resolved if  $\Delta x\ll \ell_c\ll L$ and $\Delta t\ll t_c \ll T$, where $T$ is the total simulation time.
For both equations we used a time step $\delta t=0.05$ (in 1D and 2D) and  spatial grid $\delta x=0.25$ (in 1D), $\delta x=0.65$ (in 2D), and the number of nodes for the discretization in 1D was $N_x=4096$, and in 2D in both directions was $N_x=N_y=512$. Simulations were first relaxed for a time $T_1=5\times10^6$ (1D) and $T_1=2.5\times10^5$ (2D) and later configurations were recorded every  $\Delta T=250$ (1D) and $\Delta T=50$ (2D), for a total simulation time $T_2=1\times10^7$(1D)  and  $T_2=5\times10^5$~(2D).

%\subsection{Numerical structure factor}

%Figure \ref{fig3-si} presents the non-scaled structure factor for  different values of the spatial coupling exponent $b$ in $d=1$ and $d=2$ spatial dimensions. As predicted by the scaling analysis, a decrease in $S(k)$ is obtained for $b>d$, namely $b=2$ and $b=4$ is 1D, and only for $b=4$ in 2D. 
%Complementary to the main results in the letter, the structure factor is also analyzed in detail in 1D. Again, hyperuniformity is obtained when approaching the critical point, with an exponent $\alpha=1.58 \pm 0.10$ both for  $b=2$ and $b=4$ (see Fig.~\ref{fig4-si}).

%\begin{figure}[t!]
%\includegraphics[width=0.8\columnwidth]{Fig2d-lineal.pdf}
%\includegraphics[width=0.8\columnwidth]{Fig2d.pdf}
%\caption{Static structure factor $S(k)$ obtained from the numerical solution of the model. (a) Results in 2D when $b=4$  for different values of $\nu$. The solid lines are the results of the fits with the same function. (b)  Results in 1D and 2D for different values of the spatial coupling exponent $b$. \comRS{Nestor: Cual es el valor de $\nu$?} }
%\label{fig3-si}
%\end{figure}
%

}

\showmatmethods{} % Display the Materials and Methods section

\acknow{
We thank Scott R. Waitukaitis for help during the preparation of this manuscript. 
This research is supported by Fondecyt Grants No.\ 1150393 (N.M. and J.C.S.), 1181823 (N.S.),  1180791 (R.S. and N.S.), and  3160032 (G.C.) and by the Millennium Nucleus ``Physics of active matter'' of the Millennium Scientific Initiative of the Ministry of Economy, Development and Tourism (Chile) (N.S. and R.S.).}

\showacknow{} % Display the acknowledgments section

\begin{widetext}

\newpage

\includegraphics[width=2.2\columnwidth]{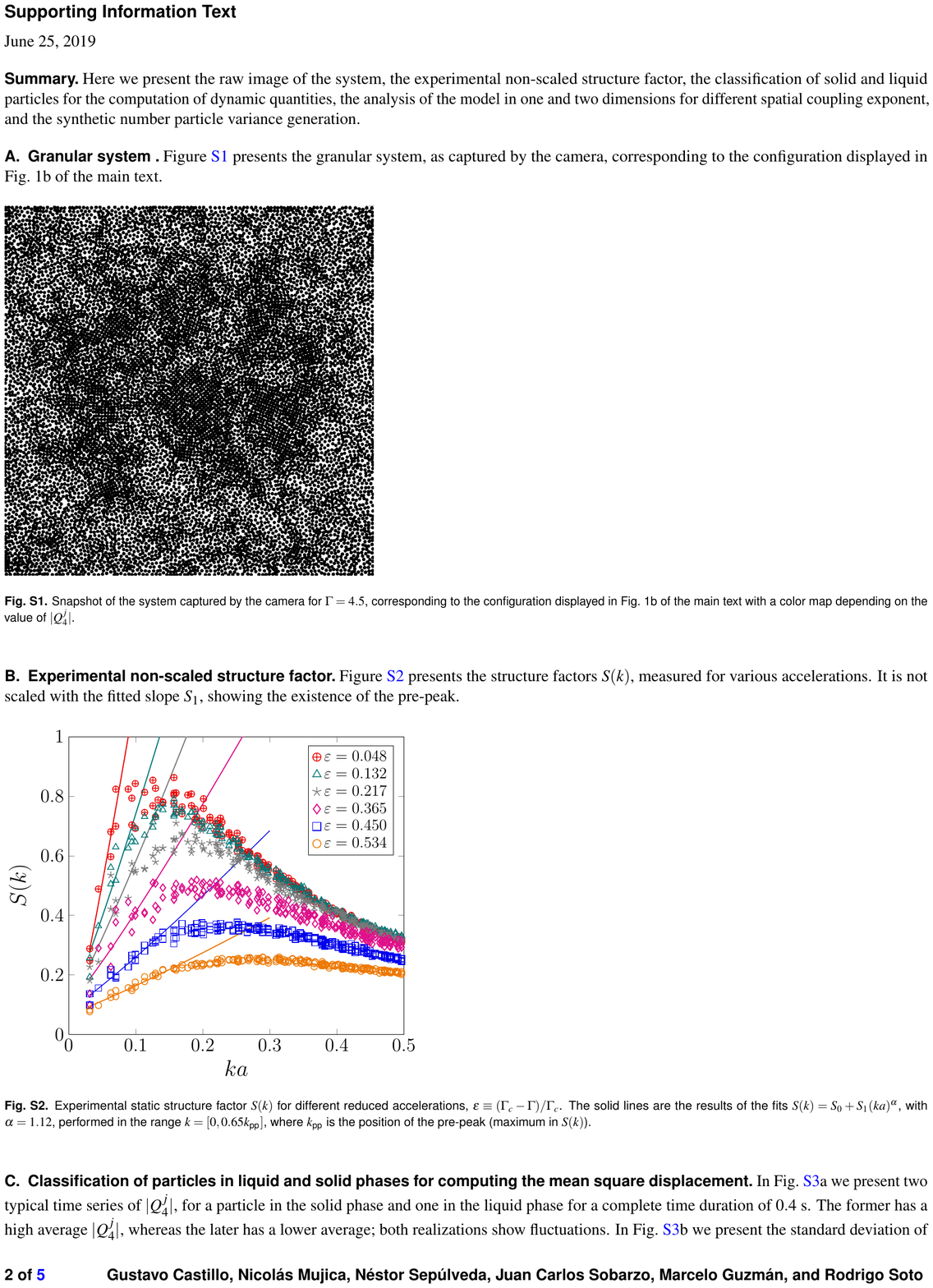}

\newpage

\includegraphics[width=2.2\columnwidth]{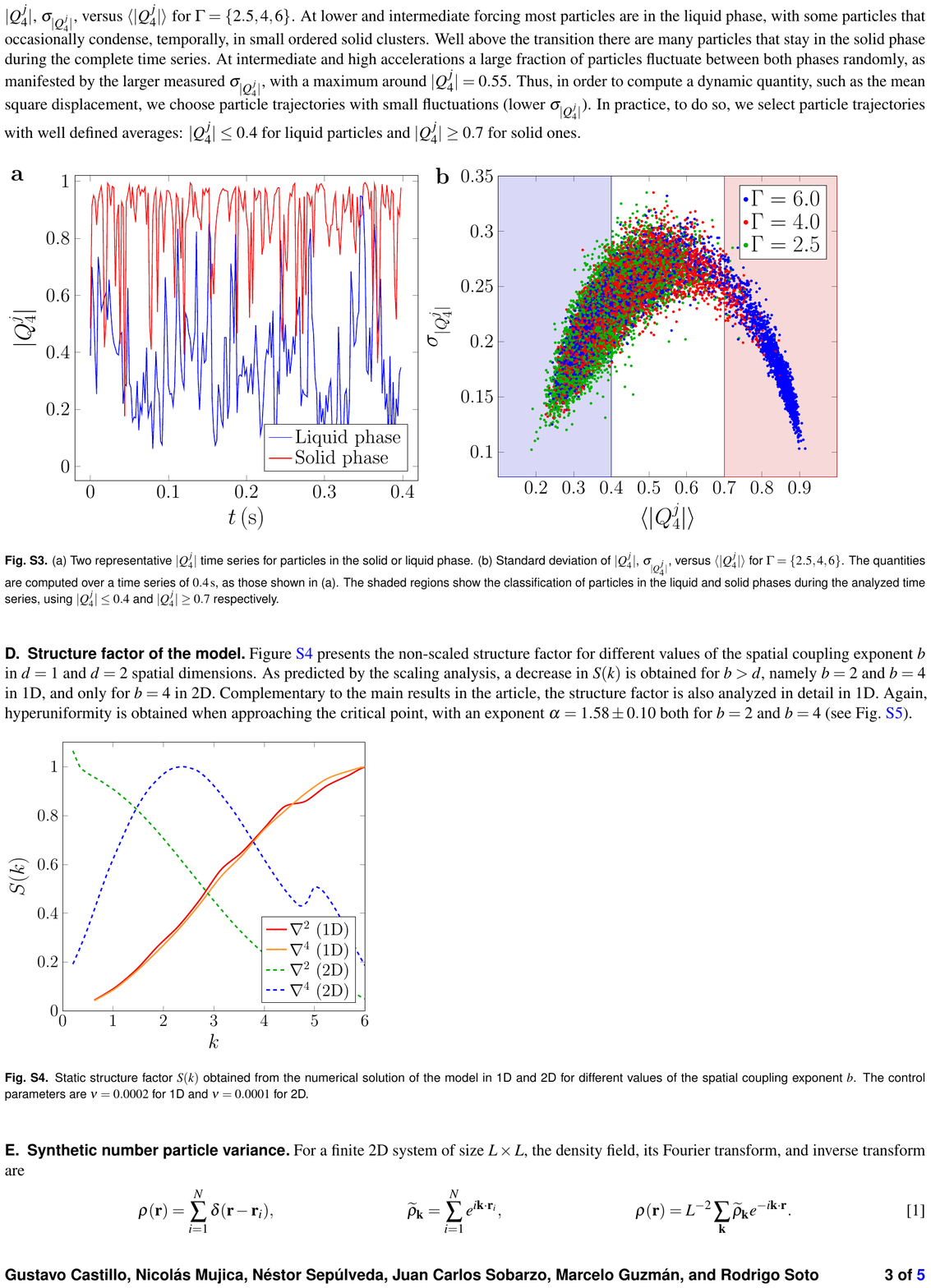}

\newpage

\includegraphics[width=2.2\columnwidth]{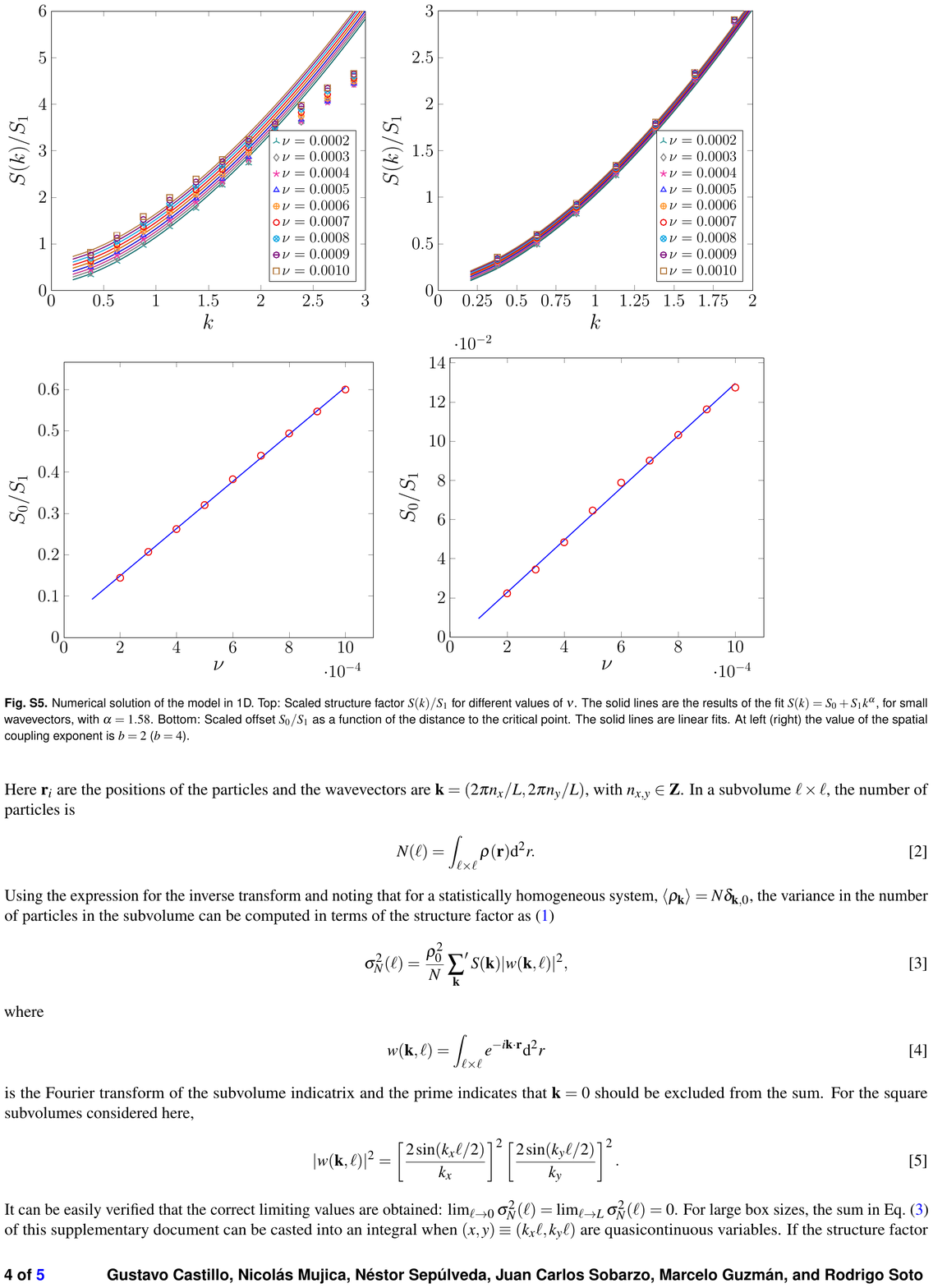}

\newpage

\includegraphics[width=2.2\columnwidth]{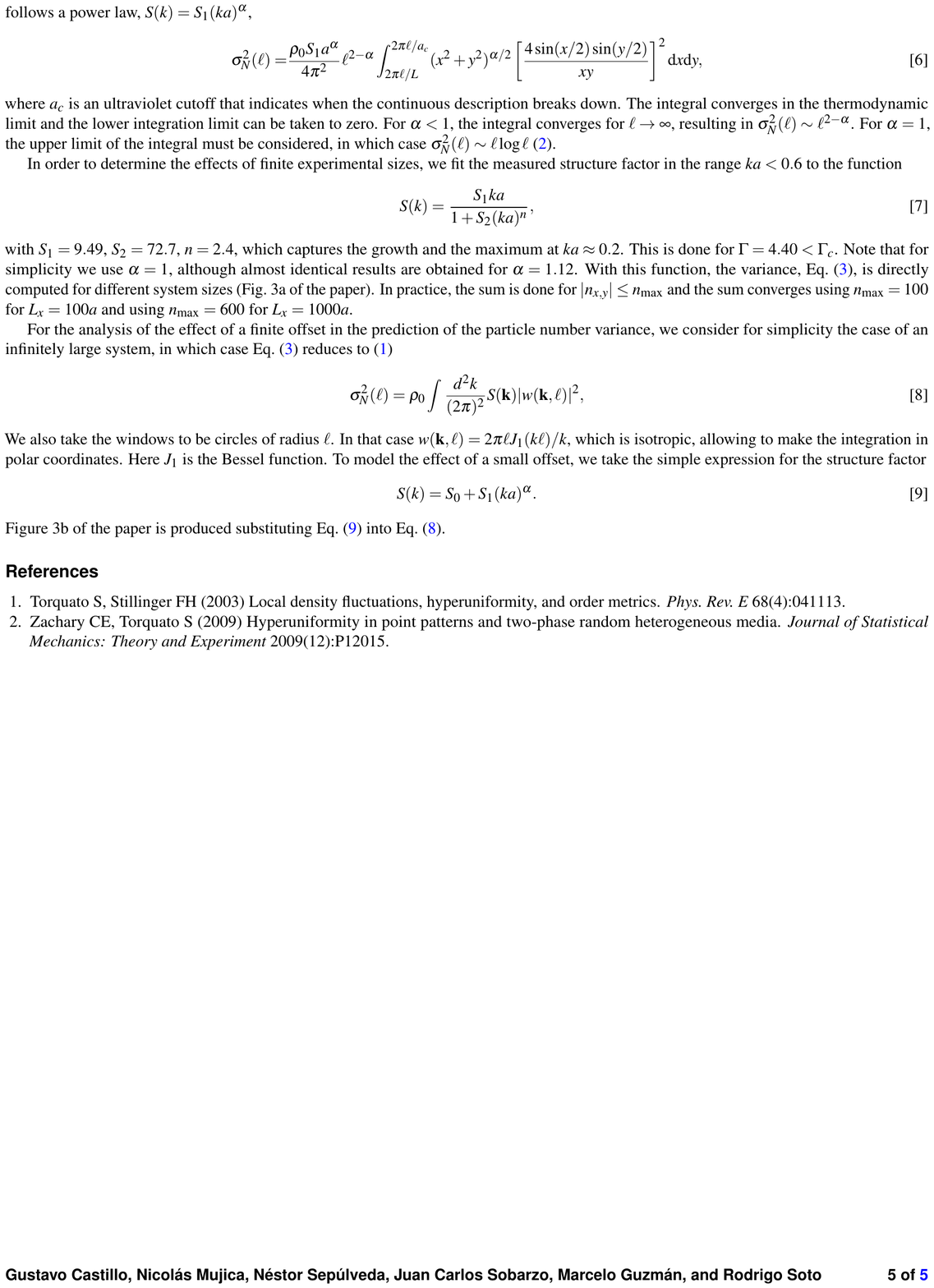}

\end{widetext}

\end{document}